\documentclass[conference, a4paper]{IEEEtran}
\IEEEoverridecommandlockouts
\usepackage{cite}
\usepackage{url}
\usepackage{amsmath,amssymb,amsfonts}
\usepackage{algorithmic}
\usepackage[pdftex]{graphicx}
\usepackage{textcomp}
\usepackage{xcolor}
\usepackage{setspace}
\usepackage{booktabs}
\usepackage{listings}

\def\BibTeX{{\rm B\kern-.05em{\sc i\kern-.025em b}\kern-.08em
    T\kern-.1667em\lower.7ex\hbox{E}\kern-.125emX}}
\begin{document}

\title{ExKaldi-RT: A Real-Time Automatic Speech Recognition Extension Toolkit of Kaldi}

\author{
    \centering
    \IEEEauthorblockN{Yu Wang, Chee Siang Leow}
    \IEEEauthorblockA{\textit{University of Yamanashi} \\
    Kofu, Yamanashi, Japan \\ \{wangyu,cheesiang\_leow\}@alps-lab.org}
    \and
    \IEEEauthorblockN{Akio Kobayashi}
    \IEEEauthorblockA{\textit{Tsukuba University of Technology} \\
    Tsukuba, Ibaraki, Japan \\ a-kobayashi@a.tsukuba-tech.ac.jp}
    \and    
    \IEEEauthorblockN{Takehito Utsuro}
    \IEEEauthorblockA{\textit{University of Tsukuba} \\
    Tsukuba, Ibaraki, Japan \\ utsuro@iit.tsukuba.ac.jp}
    \and
    \IEEEauthorblockN{Hiromitsu Nishizaki}
    \IEEEauthorblockA{\textit{University of Yamanashi} \\
    Kofu, Yamanashi, Japan \\ hnishi@yamanashi.ac.jp}      
}

\maketitle

\begin{abstract}

This paper describes the \textit{ExKaldi-RT} online automatic speech recognition (ASR) toolkit that is implemented based on the Kaldi ASR toolkit and Python language. ExKaldi-RT provides tools for building online recognition pipelines. While similar tools are available built on Kaldi, a key feature of ExKaldi-RT that it works on Python, which has an easy-to-use interface that allows online ASR system developers to develop original research, such as by applying neural network-based signal processing and by decoding model trained with deep learning frameworks. We performed benchmark experiments on the minimum LibriSpeech corpus, and it showed that ExKaldi-RT could achieve competitive ASR performance in real-time recognition.
  
\end{abstract}


\section{Introduction}

In the past few years, automatic speech recognition (ASR) has been integrated in a lot of practical products and services, including in smart home devices and intelligent driving systems. The most popular open-source ASR toolkit, Kaldi \cite{Kaldi}, provides integrated tools for building a state-of-the-art ASR system based on acoustic modeling with a Gaussian mixture model (GMM) or a deep neural network (DNN) and the construction of a decoding graph using a weighted finite-state transducer (WFST) \cite{WFST}. A DNN model has a better ability to fit distributions of acoustic features and has performed well in various ASR tasks \cite{DBN, ASR-DL}.

Deep learning (DL) shows excellent potential in ASR. A lot of research and development has been driven by some open-source DL frameworks, including TensorFlow \cite{Tensorflow2015} and PyTorch \cite{Pytorch}. Although the Kaldi ASR toolkit has tools for training a DNN model, these mainstream frameworks have enabled us to build a highly accurate DNN model for further improving the performance of ASR systems. Some existing tools, such as PyKaldi \cite{PyKaldi, PyKaldi2} and PyTorch-Kaldi \cite{PyTorch-Kaldi}, have tried to build a bridge between Kaldi and these DL frameworks. 
In our previous study, we presented the ExKaldi ASR toolkit \cite{ExKaldi}, which is one of the Kaldi wrappers in Python language. A key feature of these toolkits is their provision of an interface for computing acoustic probabilities with a DNN acoustic model trained by a DL framework followed by decoding with Kaldi's decoding program. However, these toolkits are mainly used to develop offline systems and do not have complete tools for evaluating such offline systems under actual conditions (environments). Therefore, a toolkit for the easy construction of an online ASR system is needed.

This paper introduces a new open-source toolkit named \textit{ExKaldi-RT} (Real-Time ASR Extension Toolkit of Kaldi). ExKaldi-RT is a separate part of the ExKaldi toolkit. It wraps Kaldi's functions, including online feature extraction and decoding with a lattice. Unlike the above-mentioned tools that were developed mainly for offline (not real-time) ASR, ExKaldi-RT builds an online ASR environment to enable users to apply their original recognition models and evaluate them under actual conditions (environments). Recognition models can be obtained with the above offline ASR toolkits and Kaldi. While similar online (real-time) ASR tools \cite{GSPLUG, RTAST} are available that combine Kaldi's online feature pipeline and stream management tools such as GStreamer \cite{GSTREAMER}, ExKaldi-RT implements online ASR with only Python language and is based on a DNN acoustic model trained with the DL frameworks. On the other hand, end-to-end (E2E) online ASR systems have also been developed. However, the online ASR systems using the Deep Neural Network Hidden Markov Model (DNN-HMM) are still crucial in various tasks with their improved robustness and real-time performance.

ExKaldi-RT provides integrated tools for constructing online ASR pipelines, including for recording real-time audio streams, performing voice activity detection (VAD), transmitting data when using a remote connection, computing online features, estimating probability, and decoding on the fly. Most of these jobs can be customized, which means that users can design their original algorithms for signal processing, feature extraction, socket packet compression, and N-best rescoring \cite{NBEST-RESCORE}. Thus, ExKaldi-RT can foster further research and development on online ASR.

We performed benchmark experiments on the minimum LibriSpeech \cite{LIBRISPEECH} public dataset. The experiments showed that our ExKaldi-RT toolkit could build an online ASR system with ideal ASR performance in real time. 
The contributions of our research are as follows:
\begin{itemize}
  \item Designed and implemented a stream pipeline for online ASR using Kaldi's modules, 
  \item Highly supports the customization of individual components of an online ASR pipeline, 
  \item Improved compatibility with existing DL frameworks for training well-performing acoustic models and other processes such as speech separation, and
  \item Provided the ExKaldi-RT toolkit as an open-software.
\end{itemize}

The rest of the paper is organized as follows. In Section 2, we detail the design of ExKaldi-RT. In Section 3, we present a sample code. In Section 4, we show our experiment results. In Section 5, we conclude this paper.

\section{Design of ExKaldi-RT}
\subsection{Architecture}

\begin{figure}[t]
  \centering
  \includegraphics[width=0.99\linewidth]{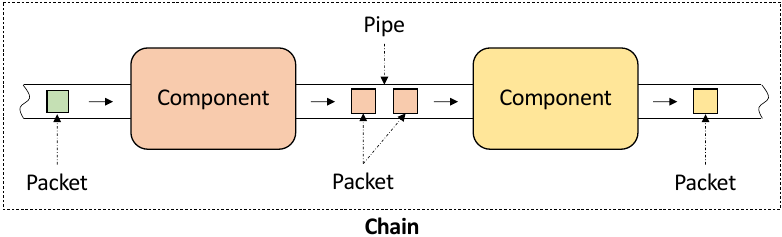} %
  \vspace*{-2mm}
  \caption{An ASR chain that connects the components of ExKaldi-RT.}
  \label{fig:units}
\end{figure}

Figure \ref{fig:units} shows the dataflow in the ASR chain composed of components and pipes of ExKaldi-RT. The data flows through the pipes are continuously processed. The ASR pipeline is basically organized with the following pieces:
\begin{itemize}
\item \textit{\bf Packet}: carries data, including digitized audio signals, acoustic features, acoustic probabilities, decoding results, and special flags such as the endpoint used to truncate the stream
\item \textit{\bf Component}: processes packets and generates new data, such as using a feature extractor and a decoder
\item \textit{\bf Pipe}: caches and transfers packets and exchanges state information between two components
\item \textit{\bf Chain}: a container that provides a high-level interface for linking and managing the ASR pipeline
\end{itemize}

In general, a component gets packets from the input side of a pipe and appends the processed results to the output side of the pipe. All the components of the chain perform the above actions simultaneously. A component monitors and modifies the state of the pipe. 
For example, when an error occurs, it propagates the error information forward and backward through the pipes.

\begin{figure}[t]
  \centering
  \includegraphics[width=0.95\linewidth]{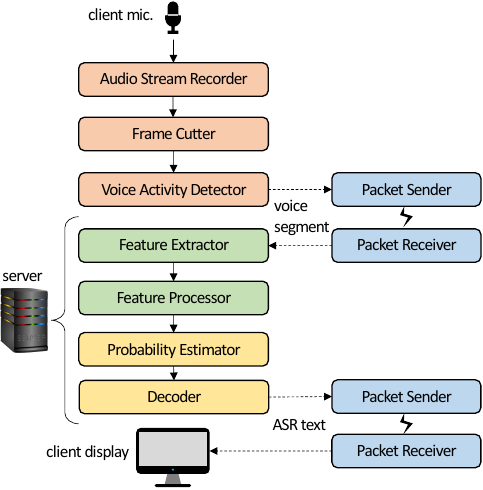}
  \vspace*{-2mm}
  \caption{An online ASR pipeline built with ExKaldi-RT.}
  \label{fig:pipeline}
\end{figure}

Figure \ref{fig:pipeline} shows a classic pipeline for online ASR built with ExKaldi-RT. Although this is a client-server mode, the pipeline can work in a series connection mode in one computer. In this pipeline, the \textit{Audio Stream Recorder} collects audio streams from a microphone connected to the client, and the \textit{Frame Cutter} cuts the stream into frames with a sliding window. The \textit{Voice Activity Detector} filters out silent audio. The components \textit{Feature Extractor} and \textit{Feature Processor} are used to compute online acoutic features. 
If a remote connection is used, the \textit{Packet Sender} and the \textit{Packet Receiver} transmit data packets between the two host computers. 
The \textit{Probability Estimator} predicts the observed probability of acoustic features, and the \textit{Decoder} outputs the ASR results to the interface of the applications.

\subsection{Voice Activity Detection}

VAD plays a remarkable role in real-time ASR. We removed long, silent audio to reduce unnecessary calculations. The VAD strategy that is used in ExKaldi-RT is executed in the following steps:
\begin{enumerate} 
  \item Detect a continuous silence sequence.
  \item For the part shorter than a threshold, keep it, and for the part longer than a threshold, the threshold discards it.
  \item Append an endpoint mark to truncate the stream, or do not.
\end{enumerate}

The silence segments are predicted using both the numerical audio stream and the acoustic feature.
Users can input some acoustic features to implement more complex VAD algorithms using a DL-based approach.

\subsection{Online Feature Extraction}

We wrapped some tools of Kaldi to implement detailed acoustic feature extraction, such as computing the fast fourier transform (FFT) and the Mel filter bank (fBank). We designed several online feature extractors to compute the spectrogram features, fBank features, Mel frequency cepstral coefficient (MFCC) features, and their combinations based on the above-mentioned tools. Besides, ExKaldi-RT supports customization of the feature extraction steps that allow users to apply novel technologies. In our experiments, we showed an example of speech separation using the magnitude spectrum and a neural network (NN) model when extracting the MFCC features. The classic feature transformation technologies, appending the differential, splicing the context features, cepstral mean and variance normalization (CMVN), and linear discriminant analysis with maximum likelihood linear transformation (LDA+MLLT), are available in ExKaldi-RT's online feature extraction. 

\subsection{Remote Transmission}

ExKaldi-RT can work on a client-server architecture. In this case, ExKaldi-RT provides tools for transmiting packets between the client and the server. A recommended way is to collect audio streams and perform VAD on the edge client. Therefore, unnecessary silent audio will not be transmitted to the feature extractor on the server. When sending packets to a remote host, the sender will add verification information to the packets' header and resend the packets if the receiver fails to verify the data. The verification information include the size of the packets, an endpoint mark, the data type, and others. In the current version, we prepared a simple function for encoding data packets without compressing them, but users can still apply their original encoding algorithms to improve the transmission efficiency. 

\subsection{Online Decoding}

ExKaldi-RT wraps the \textit{LatticeFasterDecoder} function of Kaldi and implements the real-time decoding based on a WFST. 
Instead of computing the acoustic probabilities with the \textit{NNET} model in Kaldi and other related toolkits, the decoder accepts the probabilities predicted by the acoustic probability estimator. Users can embed their original DNN-based acoustic model trained with a DL framework into the ExKaldi-RT toolkit. DNN models that invoke context or history, such as the time-delay neural network (TDNN) \cite{TDNN} and long short-term memory (LSTM) \cite{LSTM}, are also available. Besides the endpoint detection provided by Kaldi, we also recognize the endpoint flag marked by the previous components, such as the voice activity detector. If an endpoint is determined, the decoder will output N-best results. It is possible to rescore the N-best list with a ecurrent neural network (RNN)-based language model to cherry-pick the best hypothesis. 

\section{Sample Code}

The following list shows a sample code for building a typical online ASR pipeline with the ExKaldi-RT toolkit. This script can quickly deploy the series of components to complete an ASR task by performing audio recording, computing the MFCC features, predicting the acoustic probability, and decoding. A container named \textit{chain}, which links the components to each other, can easily drive the pipeline. The final results are stored in the output pipe of the \textit{chain} but are still accessible to external interface applications.

\begin{lstlisting}{python}
#Read stream from microphone and cut frames.
reader=StreamRecorder()
cutter=ElementFrameCutter()
#Compute Online MFCC features.
extractor=MfccExtractor()
processor=FeatureProcessor(FrameSlideCMVNormalizer())
#Predict probabilities with original DNN function.
extimator=AcousticEstimator()
extimator.acoustic_function=Your_Function
#Decode the probability with WFST graph.
decoder=WfstDecoder(symbolTable="words.txt",
                    silencePhones="1:2:3:4:5",
                    frameShiftSec=0.01,
                    tmodel="final.mdl",
                    graph="HCLG.fst")
#Link these components.
chain = Chain()
chain.add(reader)
chain.add(cutter)
chain.add(extractor)
chain.add(processor)
chain.add(acousticmodel)
chain.add(decoder)
#Run.
chain.start()
\end{lstlisting}

Most the components deal with data with a batch of frames in the pipeline of online ASR. Therefore, we can use a DNN model which can accept certain context frame length, such as convolutional neural network (CNN). In some components, batch data are further processed with multiple CPU threads, which is currently the bottleneck hindering the real-time factor, especially when multiple neural networks are being used.

\section{Experiments}

The experiments show that the DNN-based acoustic model that uses ExKaldi-RT performs as well as Kaldi's original model. 
Besides, external modules can be easily implemented in an online ASR pipeline built with ExKaldi-RT. To demonstrate this, therefore, the speech separation approach simply is implemented and evaluated. 

\subsection{Experimental Setup}

Our experiments used the minimum LibriSpeech corpus \cite{LIBRISPEECH}, which contains 5 hours clean data for training and 2 hours clean data for evaluation. To evaluate the online ASR performance, we simulated the real-time audio recording process by continuously reading the data stream from the files. The standard training recipe (including the DNN model training) can be found in Kaldi's examples. We used the alignments, lexicons, and decoding graph generated after the \textit{tri3b} stage of the standard recipe. In the experiments on speech separation, we used the NOISEX-92 \cite{NOISEX92} background noise dataset to synthesize the experiment data. This dataset is publicly available.
We used TensorFlow (version 2.4.0) as the DL engine to train the DNN-based model in the experiments. 
The machine configuration was as follows: CPU, Core-i7 6950X 3.0GHz; memory, 128 GB; GPU, GeForce GTX-1080Ti; and OS, Ubuntu 18.04.

\subsection{DNN Acoustic Model}

We first trained a widely-used, fully-connected NN acoustic model and embedded it into the pipeline. The \textit{word error rates} (WERs) obtained on offline and online ASR environments are shown in Table \ref{tab:off_on_com}. We compared the results of Kaldi's chain model that has an additional i-vector feature. Although we used only the MFCC feature and maximum likelihood criterion to train the NN model, thanks to the more flexible modeling approaches provided by the DL framework, we achieved an ASR performance similar to that of Kaldi's baseline. After we applyed applying this model to our online ASR system, there was a small increase of $0.16$ points in the WER. The real-time factor (RTF) showed that our online recognition system achieved a competitive real-time performance.

\begin{table}[t]
    \caption{WERs [\%] and RTF of offline and online ASR systems with the NN acoustic model.}
    \label{tab:off_on_com}
    \centering
    \vspace*{-2mm}
    \begin{tabular}{l|lll}
      \toprule
                                                  & \textbf{Offline}      & \multicolumn{2}{c}{\textbf{Online}}        \\
                                                  & \textbf{WER}          & \textbf{WER}         & \textbf{RTF}        \\
      \midrule
      Baseline                                    & $18.58$               & $18.49$              & ---                 \\
      ExKaldi-RT                                  & $19.81$               & $19.97$              & $0.57$              \\
      \bottomrule
    \end{tabular}
  \end{table}

\subsection{Acoustic Features}

The experiments considered different acoustic features, i.e., the 13 dims MFCC feature, 24 dims fBank feature, impoved 40 dims MFCC feature with LDA+MLLT transformation, and their mixture feature. Table \ref{tab:feat_com} compares the WERs and the RTFs of the online ASR systems using these features. The first three static features are processed by combining them with $\Delta$+$\Delta\Delta$ and splicing them with 10 frames of context. The mixture feature is composed of 39 dims MFCC (13 static+$\Delta$+$\Delta\Delta$), 24 dims fBank, and 40 dims LDA+MLLT, and then spliced with three frames of context. ExKaldi-RT supports the handling of multiple feature combinations. The mixed features achieved the best WER (18.56\%) at a sufficiently acceptable RTF in this set of experiments.

\begin{table}[t]
  \caption{WERs [\%] and RTFs using various acoustic features.}
  \label{tab:feat_com}
  \centering
  \vspace*{-2mm}
  \begin{tabular}{l|ll}
    \toprule
                        & \textbf{WER} & \textbf{RTF} \\
    \midrule
    MFCC                & $19.97$      & $0.57$           \\
    fBank               & $19.65$      & $0.54$           \\
    LDA+MLLT            & $19.35$      & $0.67$           \\
    Mixture             & $18.56$      & $0.75$           \\
    \bottomrule
  \end{tabular}
\end{table}

\subsection{Speech Separation}

\begin{figure}[t]
  \centering
  \includegraphics[width=0.50\linewidth]{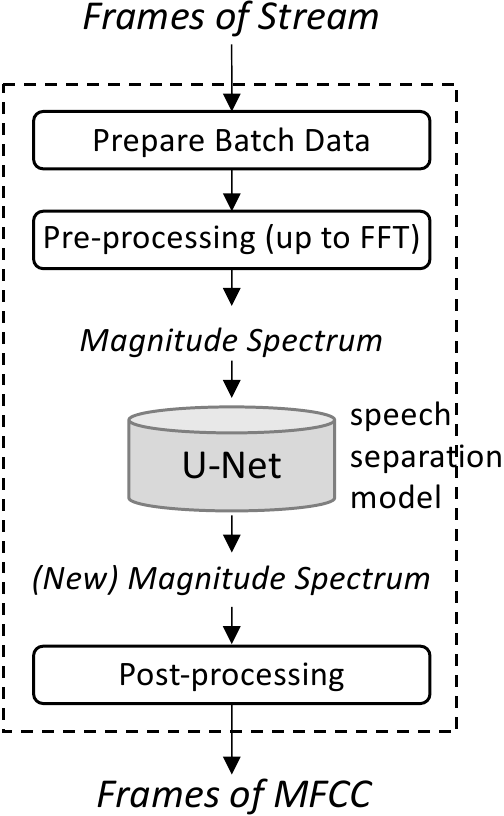}
  \vspace*{-2mm}
  \caption{Embedded the speech separation model in the ASR pipeline.}
  \label{fig:embedmodel}
\end{figure}

Speech separation with a DL-based approach has also received much interest in recent years. This set of experiments embedded a DL-based speech separation model into the feature extractor. In this study, we built a U-Net-based speech separation model \cite{DENOISE}, and we show in this paper that it works in the ASR pipeline. 
The U-Net-based speech separation model was embedded in the \textit{Feature Extractor} as shown in Figure \ref{fig:embedmodel}. We first calculated the magnitude spectrogram from the input speech signal and inputted it into the speech separation model to generate a new magnitude spectrogram with the noise removed. We simply configured a signal-to-noise ratio (SNR) of 15 dB to synthesize clean data and noise in order to generate a training and testing dataset. Table \ref{tab:denoise_com} shows the experiment results via the ASR pipeline. 
We did not collect the CMVN statistics for the new generated data in advance, so we compare the results using only sliding CMVN. When we installed the acoustic model trained on the clean dataset in a noisy environment, the WER dropped sharply. 
However, when we introduced the speech separation model, the WER was reduced by 9.49 points. We showed that the NN-based separation model can be easily used in the online ASR pipeline built by ExKaldi-RT 
and can improve the ASR performance in a more realistic noise environment.

\begin{table}[t]
  \caption{WERs [\%] and RTF using the speech separation model.}
  \label{tab:denoise_com}
  \vspace*{-2mm}
  \centering
  \begin{tabular}{l|ll}
    \toprule
                             & \textbf{WER} & \textbf{RTF}     \\
    \midrule
    clean                    & $22.66$      & $0.60$           \\
    clean+noise              & $47.58$      & $0.67$           \\
    clean+noise w/ separation & $38.09$      & $0.86$           \\
    \bottomrule
  \end{tabular}
\end{table}

\section{Conclusions}

This paper described the ExKaldi-RT toolkit, a new initiative to develop an online ASR system with Python language. ExKaldi-RT provides plug-in tools for constructing a customized online ASR pipeline by applying original algorithms that include, but are not limited to, feature extraction and probability prediction. Our experiments showed that ExKaldi-RT can achieve state-of-the-art results based on useful ASR technologies. The ExKaldi-RT toolkit is already open-source on GitHub\footnote{\url{https://github.com/wangyu09/exkaldi-rt}}. 
We hope it can be helpful for all potential researchers and developers. 
In the future, we plan to expand online decoders to support the E2E approach.

\section*{Acknowledgement}
This study was supported by JSPS KAKENHI Grant Number 20H05558, and partly by the Hoso Bunka Foundation.

\bibliographystyle{IEEEtran}
\bibliography{mybib.bib}

\begin{thebibliography}{10}
\providecommand{\url}[1]{#1}
\csname url@samestyle\endcsname
\providecommand{\newblock}{\relax}
\providecommand{\bibinfo}[2]{#2}
\providecommand{\BIBentrySTDinterwordspacing}{\spaceskip=0pt\relax}
\providecommand{\BIBentryALTinterwordstretchfactor}{4}
\providecommand{\BIBentryALTinterwordspacing}{\spaceskip=\fontdimen2\font plus
\BIBentryALTinterwordstretchfactor\fontdimen3\font minus
  \fontdimen4\font\relax}
\providecommand{\BIBforeignlanguage}[2]{{%
\expandafter\ifx\csname l@#1\endcsname\relax
\typeout{** WARNING: IEEEtran.bst: No hyphenation pattern has been}%
\typeout{** loaded for the language `#1'. Using the pattern for}%
\typeout{** the default language instead.}%
\else
\language=\csname l@#1\endcsname
\fi
#2}}
\providecommand{\BIBdecl}{\relax}
\BIBdecl

\bibitem{Kaldi}
D.~Povey, A.~Ghoshal, G.~Boulianne, L.~Burget, O.~Glembek, N.~Goel,
  M.~Hannemann, P.~Motlíček, Y.~Qian, P.~Schwarz, J.~Silovský, G.~Stemmer,
  and K.~Vesel, ``The Kaldi speech recognition toolkit,'' in \emph{Proceedings of the IEEE 2011
  Workshop on Automatic Speech Recognition and Understanding}, 2011.

\bibitem{WFST}
M.~Mohri, F.~Pereira, and M.~Riley, ``{S}peech {R}ecognition with {W}eighted {F}inite-{S}tate {T}ransducers.'' 
  Springer Handbook of Speech Processing, pp.559--584, 2008.

\bibitem{DBN}
A.~Mohamed, G.~Dahl, and G.~Hinton, ``Acoustic Modeling using Deep Belief
  Networks,'' \emph{Audio, Speech, and Language Processing, IEEE Transactions
  on}, vol.~20, pp. 14--22, 2012.

\bibitem{ASR-DL}
D.~Yu and L.~Deng, \emph{Automatic Speech Recognition: A Deep Learning
  Approach}.\hskip 1em plus 0.5em minus 0.4em\relax Springer, 2015.

\bibitem{Tensorflow2015}
M.~Abadi, P.~Barham, J.~Chen, Z.~Chen, A.~Davis, J.~Dean, M.~Devin,
  S.~Ghemawat, G.~Irving, M.~Isard, M.~Kudlur, J.~Levenberg, R.~Monga,
  S.~Moore, D.~G. Murray, B.~Steiner, P.~Tucker, V.~Vasudevan, P.~Warden,
  M.~Wicke, Y.~Yu, and X.~Zheng, ``TensorFlow: A system for large-scale machine
  learning,'' in \emph{Proceedings of the 12th USENIX Conference on Operating
  Systems Design and Implementation}, 2016, pp.265-–283.

\bibitem{Pytorch}
A.~Paszke, S.~Gross, S.~Chintala, G.~Chanan, E.~Yang, Z.~DeVito, Z.~Lin,
  A.~Desmaison, L.~Antiga, and A.~Lerer, ``Automatic differentiation in
  PyTorch,'' in \emph{Proceedings of the 31st Conference on Neural Information Processing
  Systems (NIPS 2017)}, 2017, pp. 1--4.

\bibitem{PyKaldi}
D.~Can, V.~Martinez, P.~Papadopoulos, and S.~Narayanan, ``PyKaldi: A python
  wrapper for kaldi,'' in \emph{Proceedings of 2018 IEEE International
  Conference on Acoustics, Speech and Signal Processing (ICASSP)}, 2018,
  pp. 5889--5893.

\bibitem{PyKaldi2}
L.~Lu, X.~Xiao, Z.~Chen, and Y.~Gong, ``{PyKaldi2}: Yet another speech toolkit
  based on kaldi and PyTorch,'' \emph{arXiv reprint, arXiv:1907.05955}, 2019.

\bibitem{PyTorch-Kaldi}
M.~{Ravanelli}, T.~{Parcollet}, and Y.~{Bengio}, ``The PyTorch-Kaldi speech
  recognition toolkit,'' in \emph{Proceedings of 2019 IEEE International
  Conference on Acoustics, Speech and Signal Processing (ICASSP)}, 2019, pp.
  6465--6469.

\bibitem{ExKaldi}
Y.~{Wang}, C.~S. {Leow}, H.~{Nishizaki}, A.~{Kobayashi}, and T.~{Utsuro},
  ``ExKaldi: A python-based extension tool of Kaldi,'' in \emph{Proceedings of
  2020 IEEE 9th Global Conference on Consumer Electronics (GCCE)}, 2020,
  pp. 929--932.

\bibitem{GSPLUG}
T.~Alum\"{a}e, ``Full-duplex Speech-to-text System for {Estonian}.'' IOS Press Ebooks, Vol. 268: Human Language Technologies, pp.3--10, 2014. 

\bibitem{RTAST}
C.~S. {Leow}, T.~{Hayakawa}, H.~{Nishizaki}, and N.~{Kitaoka}, ``Development of
  a low-latency and real-time automatic speech recognition system,'' in
  \emph{Proceedings of 2020 IEEE 9th Global Conference on Consumer Electronics
  (GCCE)}, 2020, pp. 925--928.

\bibitem{GSTREAMER}
``{GStreamer: open source multimedia framework},''
  \url{https://gstreamer.freedesktop.org}.

\bibitem{E2E1}
Y.~{He}, T.~N. {Sainath}, R.~{Prabhavalkar}, I.~{McGraw}, R.~{Alvarez},
  D.~{Zhao}, D.~{Rybach}, A.~{Kannan}, Y.~{Wu}, R.~{Pang}, Q.~{Liang},
  D.~{Bhatia}, Y.~{Shangguan}, B.~{Li}, G.~{Pundak}, K.~C. {Sim}, T.~{Bagby},
  S.~{Chang}, K.~{Rao}, and A.~{Gruenstein}, ``Streaming end-to-end speech
  recognition for mobile devices,'' in \emph{Proceedings of the 2019 IEEE
  International Conference on Acoustics, Speech and Signal Processing
  (ICASSP)}, 2019, pp. 6381--6385.

\bibitem{RWTH}
C.~Luscher, E.~Beck, K.~Irie, M.~Kitza, W.~Michel, A.~Zeyer, R.~Schl\"{u}ter,
  and H.~Ney, ``{RWTH ASR Systems for LibriSpeech: Hybrid vs Attention},'' in
  \emph{Proceedings of INTERSPEECH 2019}, 2019, pp. 231--235.

\bibitem{NBEST-RESCORE}
Y.~Si, Q.~Zhang, T.~Li, J.~Pan, and Y.~Yan, ``Prefix tree based N-best list
  re-scoring for recurrent neural network language model used in speech
  recognition system,'' \emph{Proceedings of INTERSPEECH 2013}, 2013, pp. 3419--3423

\bibitem{LIBRISPEECH}
V.~Panayotov, G.~Chen, D.~Povey, and S.~Khudanpur, ``LibriSpeech: An ASR corpus
  based on public domain audio books,'' in \emph{Proceedings of the 2015 IEEE
  International Conference on Acoustics, Speech and Signal Processing (ICASSP)}, 2015, pp.5206--5210.

\bibitem{TDNN}
V.~Peddinti, D.~Povey, and S.~Khudanpur, ``A time delay neural network
  architecture for efficient modeling of long temporal contexts,'' in
  \emph{Proceedings of INTERSPEECH 2015}, 2015, pp. 3214--3218.

\bibitem{LSTM}
S.~Hochreiter and J.~Schmidhuber, ``Long Short-Term Memory,'' \emph{Neural
  Computation}, vol.~9, no.~8, pp. 1735--1780, 1997.

\bibitem{NOISEX92}
A.~Varga and H.~J.~M. Steeneken, ``Assessment for automatic speech recognition
  ii: Noisex-92: A database and an experiment to study the effect of additive
  noise on speech recognition systems,'' \emph{Speech Commun.}, vol.~12, no.~3,
  pp. 247–--251, 1993.

\bibitem{DENOISE}
A.~Jansson, E.~J. Humphrey, N.~Montecchio, R.~M. Bittner, A.~Kumar, and
  T.~Weyde, ``Singing voice separation with deep U-net convolutional
  networks,'' in \emph{Proceedings of the 18th ISMIR Conference}, 2017, pp.745--751. 

\end{thebibliography}



\end{document}